\documentclass{ws-procs9x6}

\newcommand{\astroph}[2]{{astro-ph/#1}}

\begin{document}

\title{Positron Fraction  in the CMSSM\footnote{
\uppercase{P}reprint:\uppercase{IEKP-KA/2002-16}}}

\author{\underline{Wim de Boer}\footnote{\uppercase{E}-mail: wim.de.boer@cern.ch},
Markus Horn, Christian Sander}

\address{Institut f\"ur Experimentelle Kernphysik, University of Karlsruhe,
Postfach 6980, D-76128 Karlsruhe, Germany}

\author{{Dmitri Kazakov}}
\address{Bogoliubov Laboratory of Theoretical Physics, JINR,
41980 Dubna, Moscow Region, Russian Federation}

\newcommand{\tb}  {\mbox{$ \tan\beta~ $}}
%newcommand{\bsg}{$b{\to}X_s\gamma $ }
\newcommand{\besg}{$b  \to  X_s \gamma~ $}
\newcommand{\mzero}{\rm m_0}
\newcommand{\mhalf}{\rm m_{1/2}}
% add words to TeX's hyphenation exception list

\maketitle

\abstracts{
A fit to  the present cosmic ray positron fraction
can be considerably improved,
if in addition to the positron production by  nuclear interactions
in the universe the possible contribution from supersymmetric dark matter
annihilation is taken into account. 
We scan over the complete SUSY parameter space of the Constrained
Minimal Supersymmetric Model (CMSSM) and find that 
in the acceptable regions the neutralino annihilation into 
$b\overline{b}$ quark pairs is the dominant channel with hard positrons
emerging from the semileptonic decays of the B-mesons.
}

\section{Introduction} 
The cosmic ray positron fraction at momenta above 7 GeV
is difficult to describe by the background only hypothesis\cite{HEAT}.
A contribution from the annihilation of neutralinos, which are the 
leading candidates to explain the cold dark matter in the universe, 
can improve the fits considerably\cite{edsjo,kane,deboerastro}. 
The neutralinos are the Lightest Supersymmetric Particles (LSP) 
in  supersymmetric extensions of the Standard Model, which are stable, 
if R-parity is conserved.  This new multiplicative quantum number 
for the supersymmetric 
partners of the Standard Model (SM) particles is needed to prevent 
proton decay and simultaneously prevents the LSP a) to decay into the lighter 
SM particles and b) can only interact with normal matter by producing 
additional supersymmetric particles. The cross sections for the latter 
are typically of the order of the weak cross sections, so the LSP is 
``neutrinolike'', i.e. it would form halos around the galaxies 
and consequently, it is an excellent candidate for dark matter. 
 
In addition to being of interest for cosmo\-logy, supersymmetry solves 
also many outstanding problems in particle physics, among them\cite{rev}: 
1) it provides a  unification of the strong and electroweak 
forces, thus being a prototype theory for a Grand Unified Theory (GUT) 
2) it predicts spontaneous electroweak  symmetry breaking (EWSB) 
by radiative corrections through the heavy top quark, thus 
providing a relation between the GUT scale, the electroweak scale 
and the top mass, which is perfectly fullfilled 
3) it includes gravity 
4) it cancels the quadratic divergencies in the Higgs mass in the SM 
5) the lightest Higgs mass can be calculated to be below 125 GeV in perfect 
    agreement with electroweak precision data, which prefer 
    indeed a light Higgs mass. 
%%%%% 
Therefore it is interesting to study the positron fraction from neutralino 
annihilation in the reduced region of SUSY parameter 
space, where these constraints are satisfied and compare it with
 data, as  will be done in the  
next section.
It should be noted that the positrons from  the annihilation of heavy
particles leads to a much harder energy spectrum than the positrons expected
from nuclear interactions, thus leading to a significant change
in the shape of the spectrum. In addition, in the positron fraction,
defined as $e^+/(e^- +e^+)$, many systematic uncertainties related
to fluctuations in the nuclear densities and propagation in the universe,
cancel.
\begin{figure} 
\begin{center} 
\includegraphics [width=0.49\textwidth,clip]{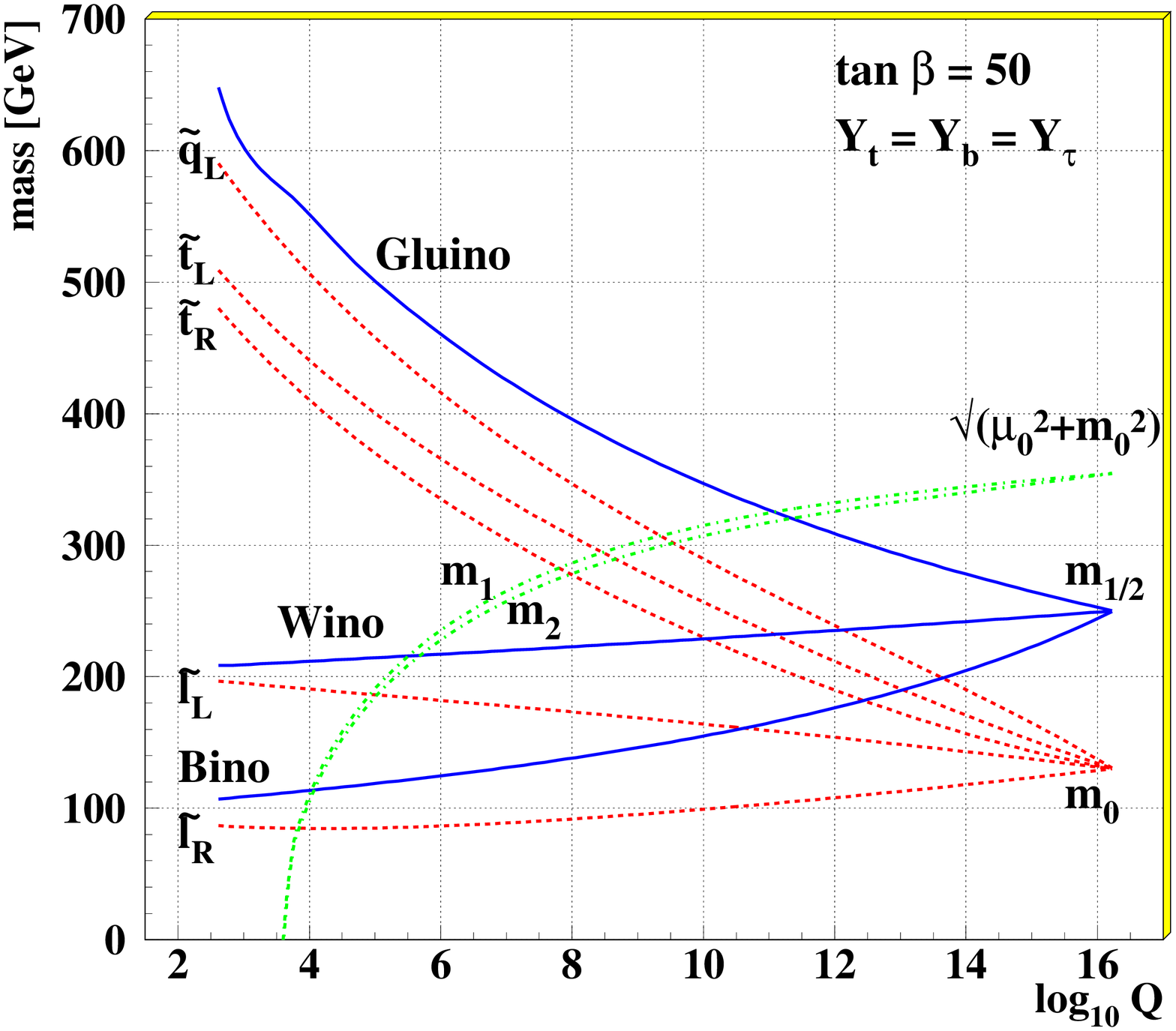}
\includegraphics [width=0.49\textwidth,clip]{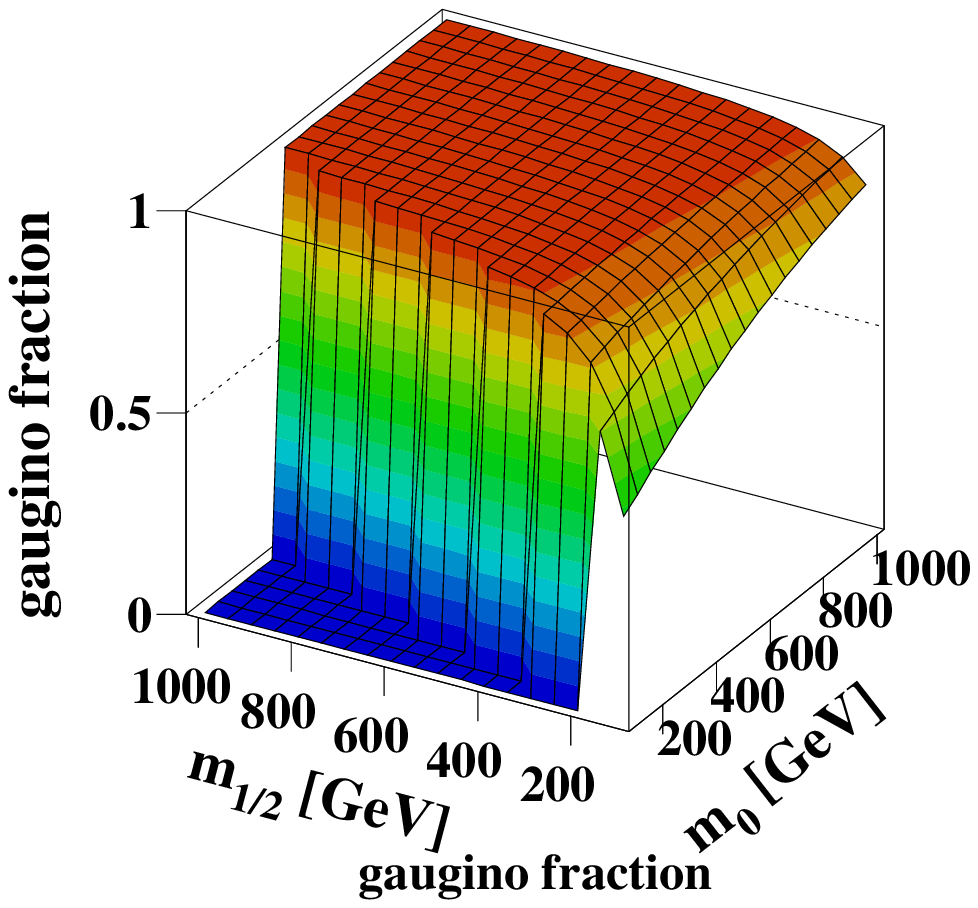} 
\caption[]{\label{mass} \it 
Left: The running of the masses between the GUT scale and the electroweak
scale for \tb=35.
Right: The gaugino fraction 
as function of $m_0$ and $m_{1/2}$ for 
$\tan\beta=35$. For lower va\-lues of \tb the gaugino fraction is even
closer to one. The region with gaugino fraction zero is excluded, since the
LSP would be charged, which is excluded because the dark matter is neutral. 
} 
\end{center} 
\end{figure} 
\section{Neutralino Annihilation in the  CMSSM} 
\begin{figure} 
\begin{center} 
\includegraphics [width=0.45\textwidth,clip]{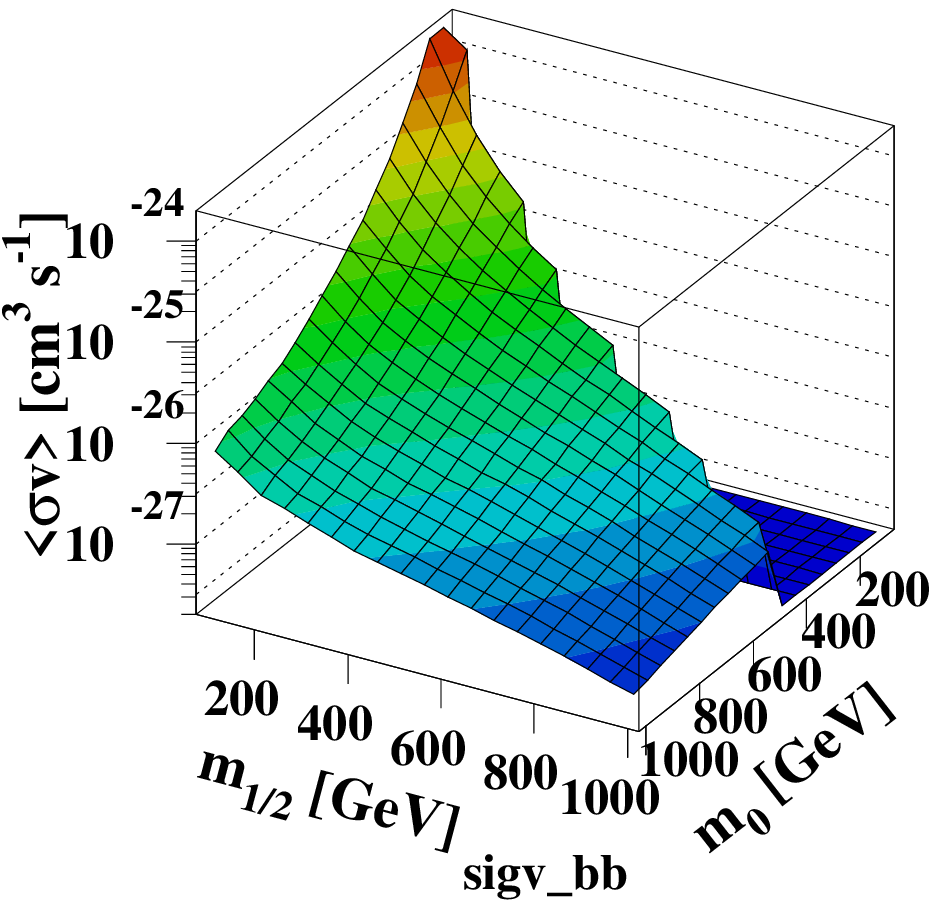} 
\includegraphics [width=0.45\textwidth,clip]{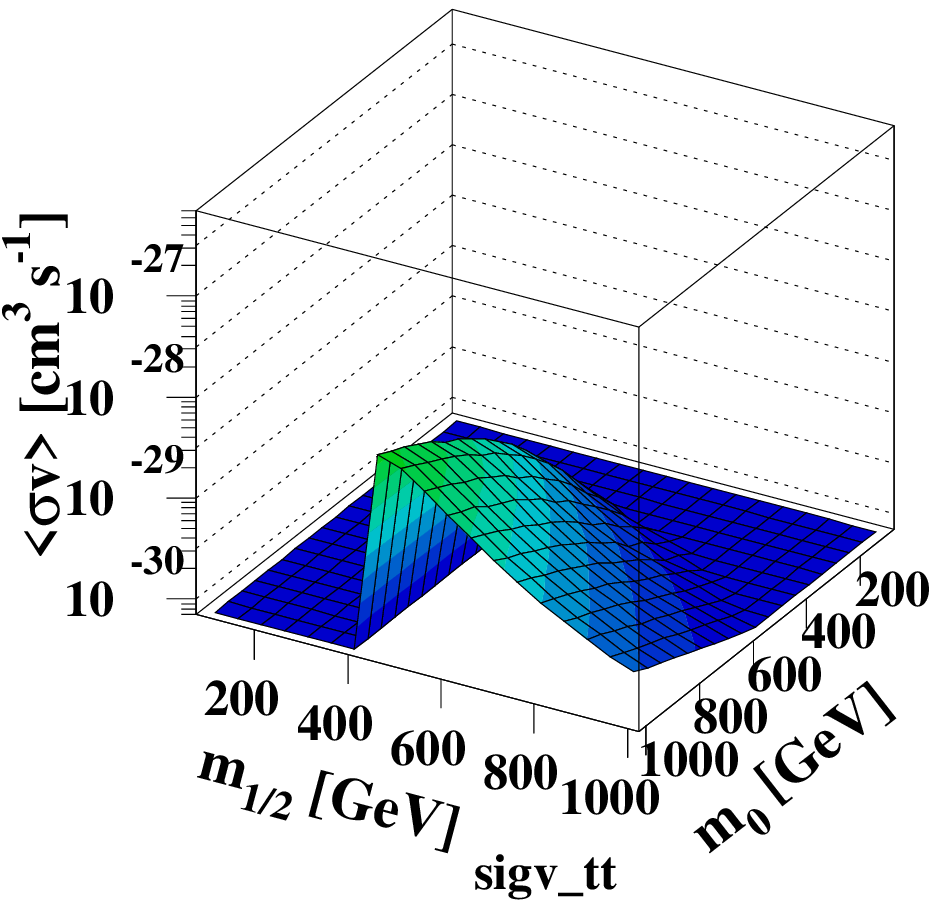} 
\includegraphics [width=0.45\textwidth,clip]{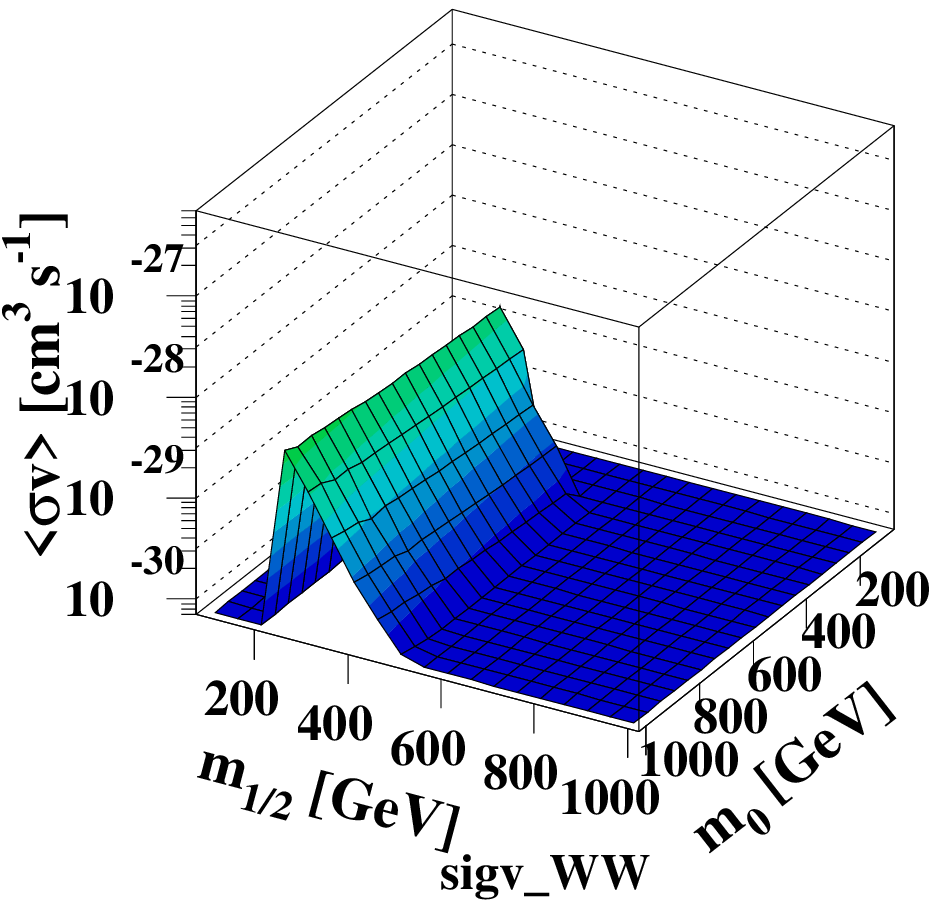}
\includegraphics [width=0.45\textwidth,clip]{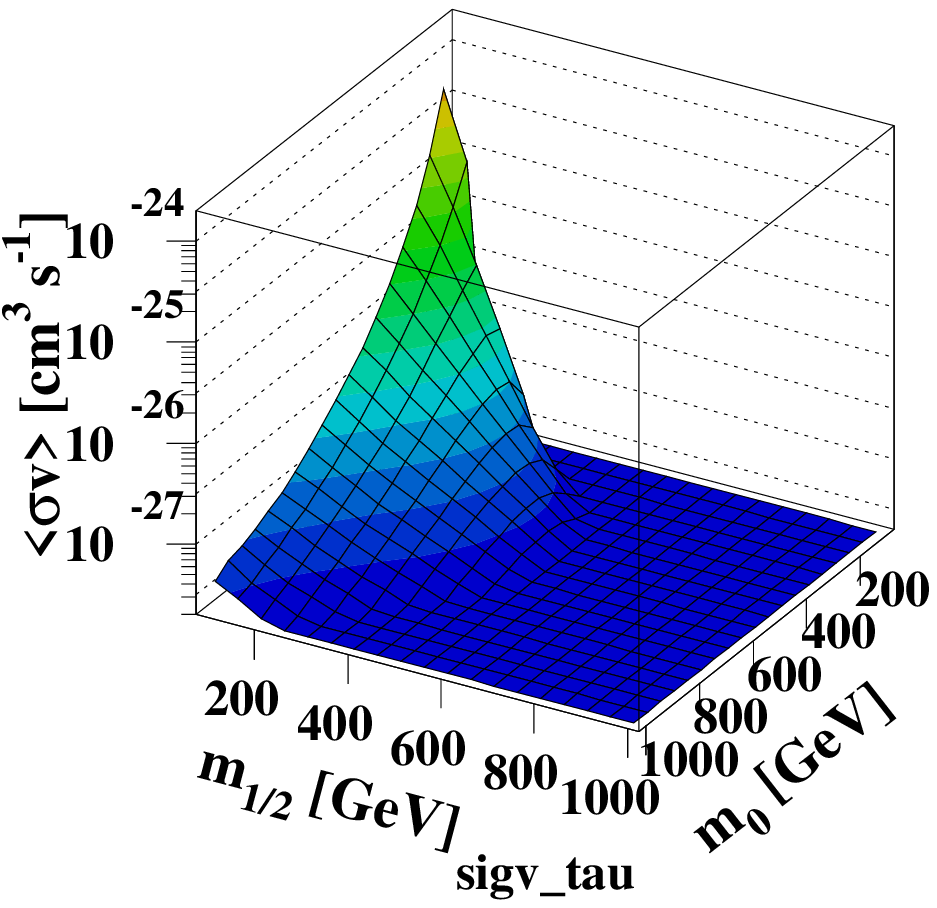} 
\caption[]{\label{sigmav} \it 
The thermally averaged cross section times velocity for 
neutralino annihilation   
as function of $m_0$ and $m_{1/2}$ for  different final states  
($b\overline{b},~ t\overline{t},~ W^+W^-,~ \tau^+\tau^-)$ at \tb=50.
The neutralino mass equals $\approx 0.4 m_{1/2}$ in the CMSSM,
so in the plots the neutralino varies from 
40 to 400 GeV along the front axis. 
The $b\overline{b}$ final state grows with $\approx \tan^2\beta$
and dominates for \tb above 5.
Note  the different vertical scales.
} 
\end{center} 
\end{figure} 
\begin{figure} 
\begin{center} 
\includegraphics [width=0.45\textwidth,clip]{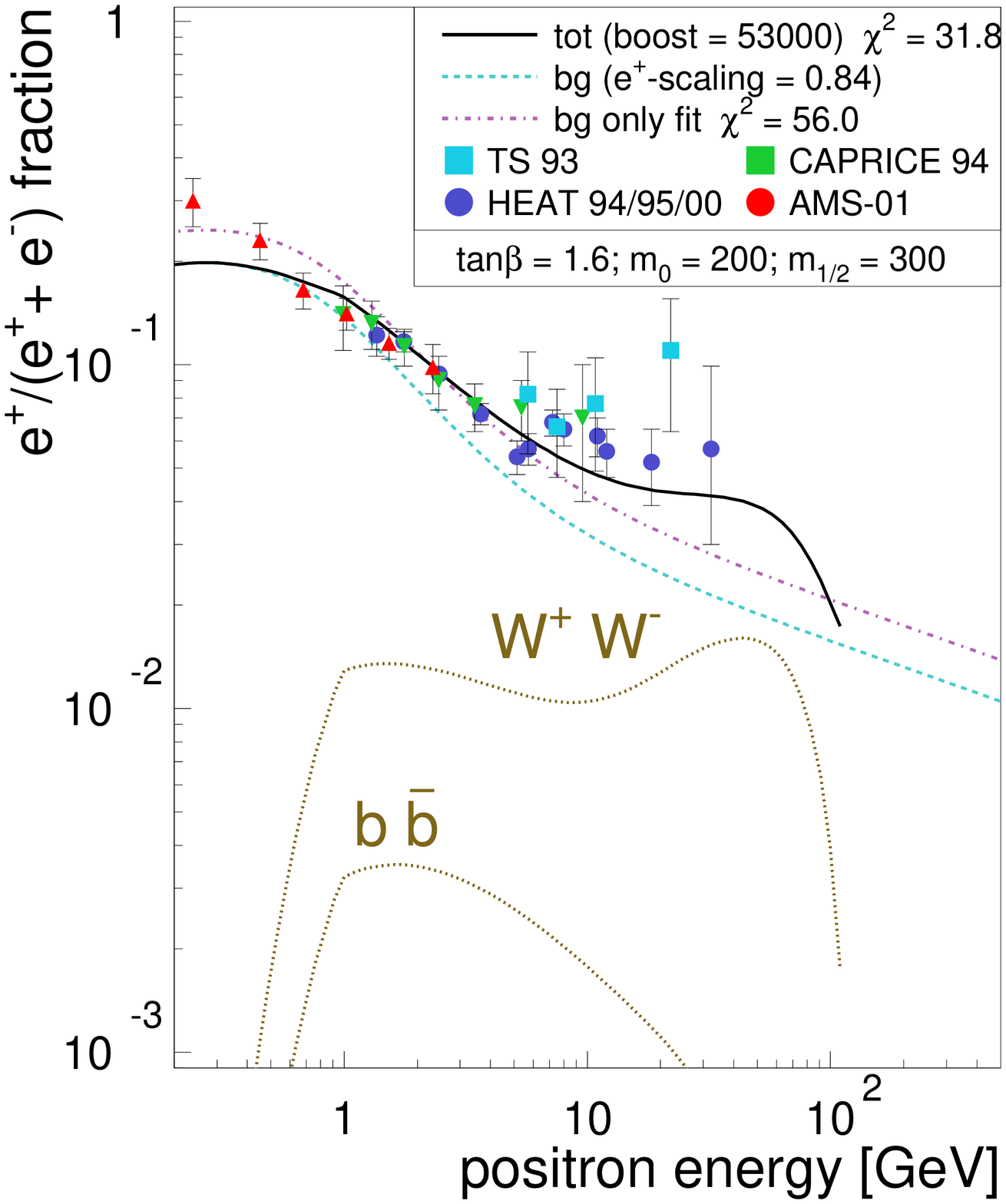} 
\includegraphics [width=0.45\textwidth,clip]{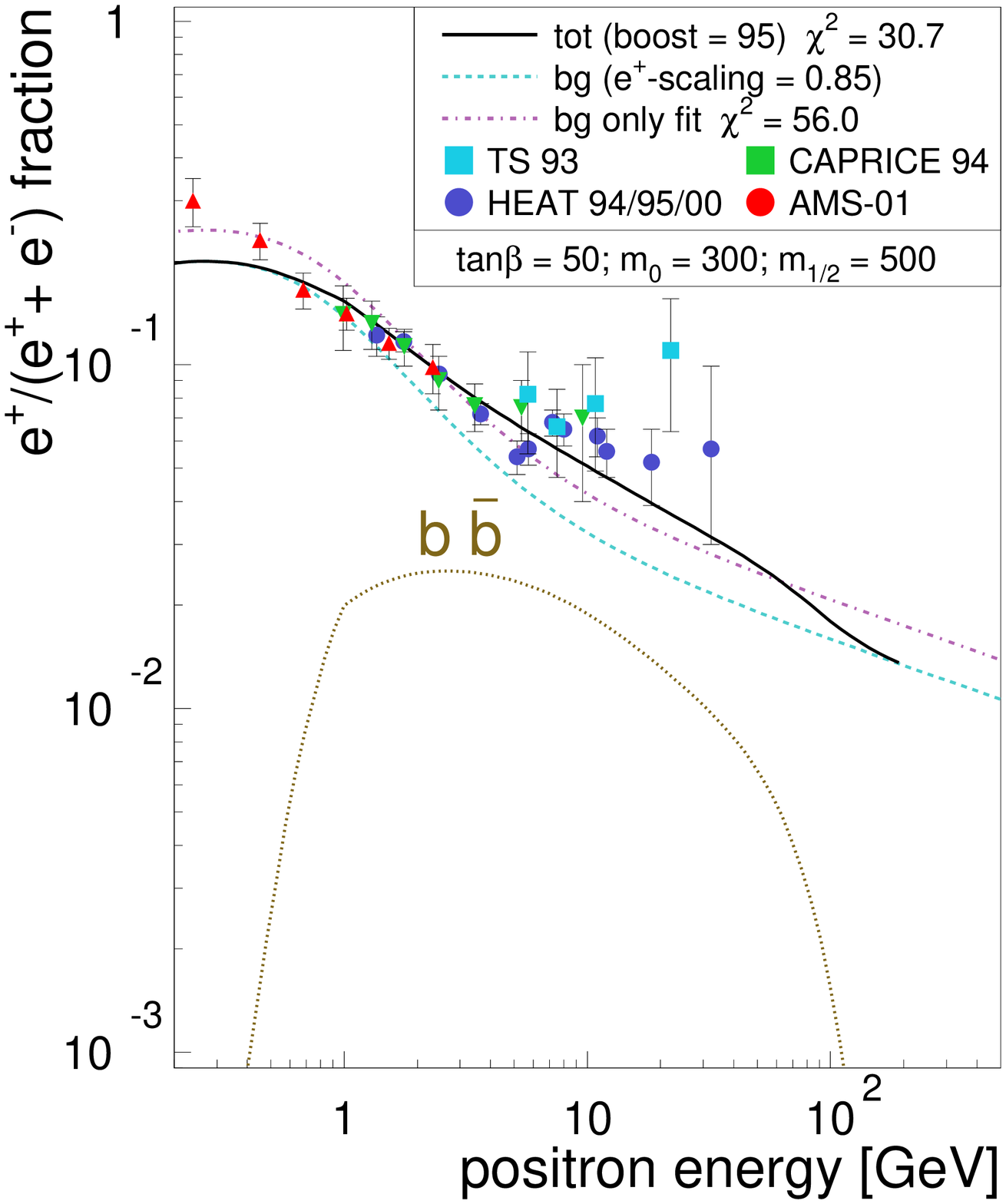} 
\caption{\label{fit} \it 
Fits to the data for \tb=1.6 and \tb=50. In the first case
 $W^+W^-$ is the  dominant annihilation 
channel, but in the second case
 $b\overline{b}$ dominates. In both cases the $\chi^2$ reduces from 
56 for the background only fit  for 27 data points to around 31, if neutralino
annihilation is included.
} 
\end{center} 
\end{figure} 
\begin{figure}
\begin{center} 
\includegraphics [width=0.4\textwidth,clip]{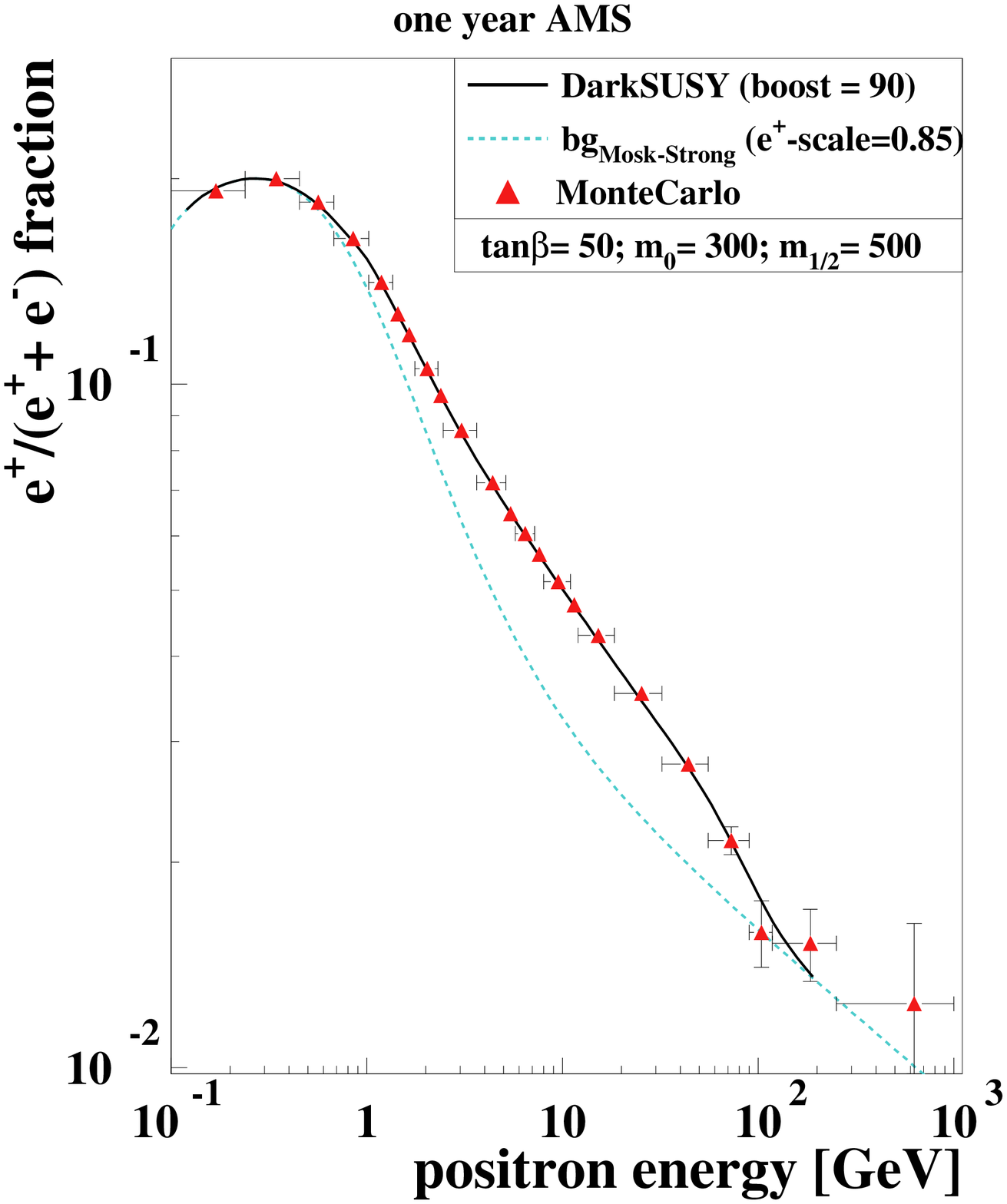}
\includegraphics [width=0.5\textwidth,clip]{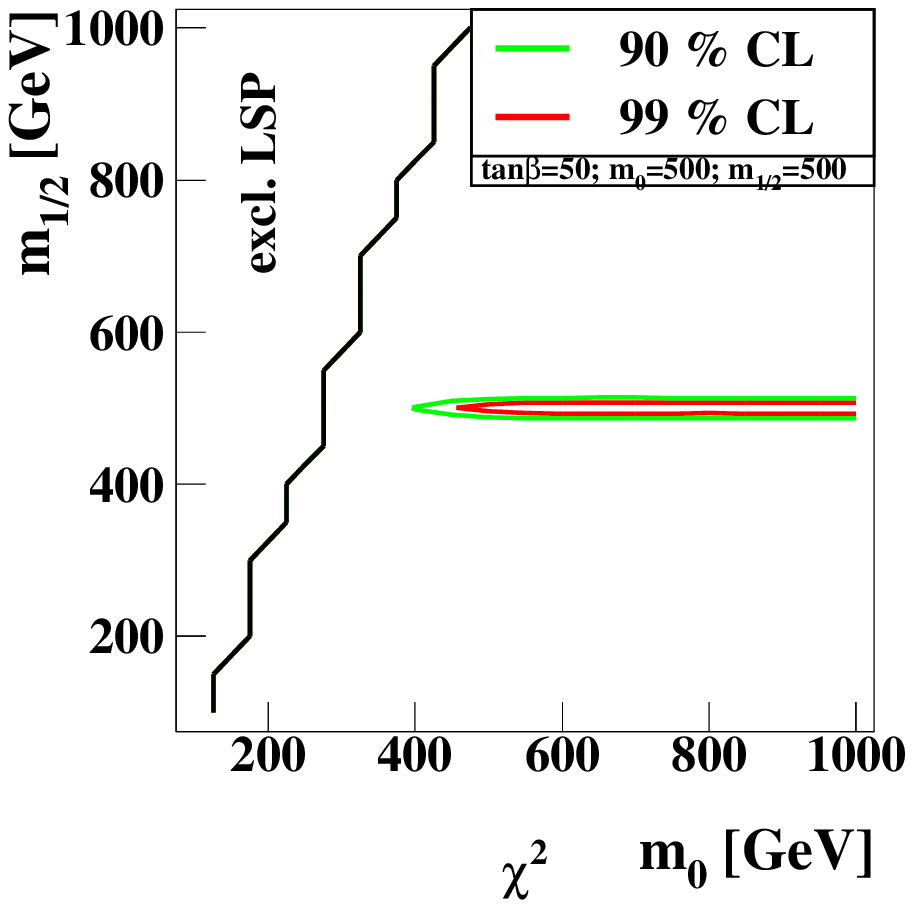} 
\caption[]{\label{fitams} \it 
Expected statistics after one year of data taking with AMS-02
for a neutralino of 200 GeV and assuming a boost factor of 90 required
by the best fit to the present  data. On the r.h.s. the $\chi^2$ contour
of the CMSSM fit to the Monte Carlo data on the left.
Note the precise determination of $m_{1/2}$, which implies a precise
determination of the LSP mass. The region labeled ``excl. LSP''
is excluded because the LSP is the charged stau; a charged stable
particle is not a dark matter candidate.
%For $m_{1/2}$ above 200 GeV the $\chi
} 
\end{center} 
\end{figure} 
 
In the Constrained Minimal Supersymmetric 
Model (CMSSM) with supergravity mediated breaking terms all sparticle 
masses are related by the usual GUT scale boundary conditions of  a 
 common mass $m_0$  for the  squarks and sleptons 
 and a common mass $m_{1/2}$ for the gauginos. 
The  parameter space, 
where all low energy constraints are satisfied, 
%both $a_\mu$ and \besg are within errors consistent with the data, 
is most easily determined by a global statistical analysis, in which 
the GUT scale parameters
%\footnote{The supergravity inspired parameters
%are the GUT scale $M_{GUT}$, the common gauge coupling $\alpha_{GUT}$,
%the common mass scales $m_0, ~m_{1/2} $ for the spin 0 and spin 1/2
%sparticles, the trilinear couplings $A_0$ and Yukawa couplings
%$Y_0$ at the GUT scale of the third generation of fermions,
% the Higgs mixing parameter $\mu$ and \tb, the ratio of vacuum
%expectation values  of the
%neutral Higgs doublets.  }
 are constrained to the low energy data by 
a $\chi^2$ minimization.
 For details we refer to previous 
publications\cite{ZP}. 
%We use the full NLO renormalization group equations\cite{rev} to calculate 
%the low energy values of the gauge and Yukawa couplings and the one-loop 
%RGE equations for the sparticle masses with decoupling of the 
%contribution to the running of the coupling constants at threshold. 
The l.h.s. of Fig. \ref{mass} shows a typical running from the common masses
at the GUT scale to low energies. The squarks and gluinos get 
a higher mass than the sleptons due to the gluonic contributions
from the strong interactions. 
If $m_{1/2}$ is not strongly
above $m_0$, the lightest mass is the supersymmetric partner of 
U(1) gauge boson, the bino, which mixes with the $W^3$ boson
and spin 1/2 higgsinos to neutralinos. 
The low energy value of the LSP is roughly 0.4 times its starting
value at the GUT scale, i.e. 0.4$m_{1/2}$.
The gaugino fraction of the LSP is defined as
$(N_{i,1}^2+N_{i,2}^2)$, where the  coefficients determine the   
neutralino mixing
$\tilde{\chi}_i^0=N_{i,1}\tilde{B}+N_{i,2}\tilde{W}^3+N_{i,3}\tilde{H}^0_1+N_{i,4}\tilde{H}^0_2$.
As shown on the r.h.s of  Fig. \ref{mass} the gaugino fraction is close to one
for LSP masses above 100 GeV, i.e. $m_{1/2}>250$ GeV.
The gaugino fraction is important, since the neutralino
properties are quite different for a pure gaugino or pure Higgsino.

Neutralino annihilation  can occur through Z- and Higgs exchanges in the 
s-channel and sfermion exchange in the t-channel.
This annihilation in the halo of the galaxies will produce antimatter 
at high momenta, thus  anomalies  in the  spectra of
positrons and antiprotons provide an excellent signal
for dark matter annihilation.
The neutralino is a spin 1/2 Majorano particle, 
so it obeys the Pauli principle, which
implies an asymmetric wave function or spins antiparallel for
annihilation at rest.
%This results in a p-wave
%amplitude to the fermion-antifermion final states,
%which is proportional to the mass of the fermion in the final state.
%Therefore  heavy final states  are enhanced at low momenta.
Final states couple most strongly with total spin 1, so light particles
are ``helicity-suppressed''.
The cross section for $b\overline{b}$ grows quickly with \tb and dominates
for large \tb, as shown in Fig. \ref{sigmav}.

%The boost factors needed
%for the best fit to describe the HEAT data at large momenta
%are correspondingly lower.%, as shown in Fig. \ref{boost}.
%These boost factors were used in the fit as an arbitrary   normalization,
%since the dark matter is not expected to be homogeneous,
%but shows some clumpiness due to gravitational interactions.
%Since the annihilation rate is proportional to the square of the dark
%matter density, the clumpiness can enhance the annihilation rate.
%by several orders of magnitude.
%Alternative ways to increase the annihilation cross section
%include vacuum energy\cite{quin}.

%The cross section calculations 
%were done with CalcHEP\cite{calchep} and DarkSUSY\cite{darksusy}, which 
%were found to  agree within a factor of two for the $t\overline{t}$  
%final state and considerably better for other final states. 

Fig. \ref{fit} shows the fit to the data  for different 
regions of parameter space with different main annihilation 
channels. The fits were done with DarkSusy and follow the same 
principle as the ones from Ref. \cite{edsjo}, i.e. the background 
is left free within a normalization error of about 20\%  
and the signal can be enhanced by a boost factor, which takes
into account the uncertainty of the relic density in our galaxy
due to possible clumpiness of the dark matter.

It should be noted, that the fits with $W^+W^-$  and $t\overline{t}$
final states are excluded by the present electroweak data,
which requires \tb above 5, in which case the $b\overline{b}$
final state dominates.
%The positron spectrum from b-quark decays 
%can fit the present data as well as the harder decays from W-pairs, 
%if the neutralinos are somewhat heavier, as shown 
%by  Fig. \ref{fit}: 
%the fit with b-quark final states
%yields a $\chi^2/d.o.f$ of 26/16, which is not much worse than
%the one for $W^+W^-$ final states (24/16).
%These values are considerably better than the background only fit,
%which yields 48/16. Here we used the background from 
%Moskalenko and Strong\cite{ms}, as implemented in DarkSUSY.
%Since the cross sections for $b\overline{b}$ final states
% are at least 
%an order of magnitude above the cross sections for 
%$t\overline{t}$ and $W^+W^-$ at large \tb,
%the needed boost factor for the best fit  
%is correspondingly lower, especially for lower values of $m_0$, as shown 
%before in Fig. \ref{boost}. 
%The region with large cross section and correspondingly lower
%boost  is also the region with  a relic density parameter
%between 0.1 and 0.3 of the critical density, as  
%in Fig. \ref{relic}. A  range between 0.1 and 0.3 is
%preferred by  the determination of the cosmological parameters
%from the red shift of distant supernovae and the
%acoustic peak in the microwave background\cite{sn}. 
% 
%The fits were repeated for all values of $m_0$ and $m_{1/2}$. The
%resulting $\chi^2$ values are plotted in Fig. \ref{chi2}.
%One observes a fast decrease in $\chi^2$ for values of $m_{1/2}$
%above 230 GeV, i.e. for LSP masses above $\approx$  100 GeV. 
%
Unfortunately the data are not precise enough to prefer a certain
value of the LSP mass. However, the present balloon experiments correspond
to only a few days of data taking. With future experiments, like
PAMELA\cite{pamela} on a russian satellite or AMS-02\cite{ams02} on the ISS
(International Space Station),
one will take data for several years, thus being able to decide
if the shape of the positron fraction originates from two components:
a hard component from neutralino annihilation and a softer component
from nuclear interactions. Taking the $\chi^2$
difference between the fit with a single
component and two components as indication of neutralino annihilation,
one can predict what the data after one year of data taking with
 AMS-02\cite{ams02}
with an acceptance of more than $0.04 ~m^2~sr$ will look like.
This is indicated by
the points in Fig. \ref{fitams}.
Here we assumed a boost factor of 90 as given
by the fit to the present data.
Clearly,
 the positron spectra can provide strong evidence for neutralino annihilation
and the data will be accurate enough to precisely determine the
neutralino mass, as shown by the r.h.s. of Fig. \ref{fitams}.
We like to thank Drs. L.  Bergstr\"om, J. Edsj\"o and P. Ullio  for helpful  
discussions.

\end{document}